\documentclass[12pt]{amsart}
\usepackage{subcaption,fullpage}
\usepackage{amsthm,amsmath,amssymb,graphics,tikz,url,hyperref}
\usepackage{lmodern,microtype}

\usepackage{environ}
\usepackage{enumitem}
\usepackage{wrapfig}
\usepackage{mdframed}
\usepackage[textsize=tiny,textwidth=.8in]{todonotes}
\usepackage[utf8]{inputenc}
\usepackage[backend=biber,style=alphabetic,maxnames=8, sorting=ynt,doi=true,isbn=false,url=false,eprint=false]{biblatex}
\newtheorem{theorem}{Theorem}
\newtheorem{observation}[theorem]{Observation}
\newtheorem{corollary}[theorem]{Corollary}
\newtheorem{lemma}[theorem]{Lemma}

\definecolor{defblue}{rgb}{0.1,0.1,0.7}
\newcommand{\defi}[1]{\textcolor{defblue}{\emph{#1}}}

\newcommand{\cF}{\mathcal{F}}
\newcommand{\cI}{\mathcal{I}}
\newcommand{\cO}{\mathcal{O}}

\newcommand{\RR}{\mathbb{R}} 

\DeclareMathOperator{\conv}{conv}
\DeclareMathOperator{\argmin}{argmin}

\newcommand{\OPT}{\textrm{OPT}}

\newcommand{\tlow}{T^+}
\newcommand{\thigh}{T^-}

\makeatletter
\newcommand{\problemtitle}[1]{\gdef\@problemtitle{#1}}
\newcommand{\probleminput}[1]{\gdef\@probleminput{#1}}
\newcommand{\problemquestion}[1]{\gdef\@problemquestion{#1}}
\NewEnviron{problem}{
  \problemtitle{}\probleminput{}\problemquestion{}
  \BODY
  \par\addvspace{.5\baselineskip}
  \noindent
  \begin{tabularx}{\textwidth}{@{\hspace{\parindent}} l X c}
    \multicolumn{2}{@{\hspace{\parindent}}l}{{\@problemtitle}} \\
    \textbf{Given:} & \@probleminput \\
    \textbf{Compute:} & \@problemquestion
  \end{tabularx}
  \par\addvspace{.5\baselineskip}
}
\makeatother 

\surroundwithmdframed[linecolor=black,linewidth=1pt,skipabove=3mm,skipbelow=0mm,leftmargin=0mm,rightmargin=0mm,innerleftmargin=3mm,innerrightmargin=3mm,innertopmargin=3mm,innerbottommargin=1mm]{algo}

\newenvironment{algo}[1]%
{\refstepcounter{theorem}\textbf{Algorithm \thetheorem} \emph{(#1)}.}{\vspace{1mm}}

\title{Set Selection with Uncertain Weights: Non-Adaptive Queries and Thresholds}

\author[C. Dürr]{Christoph Dürr}
\address[C. Dürr]{Sorbonne Université, CNRS, Laboratoire d'informatique de Paris 6, LIP6, Paris, France. Part of the work was done while C.D.\ was affiliated with Universidad de Chile, CNRS, CMM, Santiago, Chile.}
\author[A. Merino]{Arturo Merino}
\address[A. Merino]{Saarland University, Computer Science department, Saarbrücken, Germany}
\author[J. A. Soto]{José A. Soto}
\address[J. A. Soto]{Department of Mathematical Engineering, Universidad de Chile, Chile}
\author[J. Verschae]{José Verschae}
\address[J. Verschae]{Pontificia Universidad Católica, Institute for Mathematical and Computational Engineering, Faculty of Mathematics and School of Engineering, Chile}

\begin{document}
\begin{abstract}
    We study set selection problems where the weights are uncertain.
	Instead of its exact weight, only an uncertainty interval containing its true weight is available for each element.
	In some cases, some solutions are universally optimal; i.e., they are optimal for every weight that lies within the uncertainty intervals. 
	However, it may be that no universal optimal solution exists, unless we are revealed additional information on the precise values of some elements. 

	In the \emph{minimum cost admissible query} problem, we are tasked to (non-adaptively) find a minimum-cost subset of elements that, no matter how they are revealed, guarantee the existence of a universally optimal solution.
	This belongs to the setting of explorable uncertainty and while there is a significant body of work in the adaptive setting, non-adaptive versions, such as the one in this paper, are far-less understood.

	We introduce \emph{thresholds under uncertainty} to analyze problems of minimum cost admissible queries.
    Roughly speaking, for every element $e$, there is a threshold for its weight, below which $e$ is included in all optimal solutions and a second threshold above which $e$ is excluded from all optimal solutions.
    
    We show that computing thresholds and finding minimum cost admissible queries are essentially equivalent problems.
	Thus, the analysis of the minimum admissible query problem reduces to the problem of computing thresholds. 

	We provide efficient algorithms for computing thresholds in the settings of minimum spanning trees, matroids, and matchings in trees; and NP-hardness results in the settings of $s$-$t$ shortest paths and bipartite matching.
	By making use of the equivalence between the two problems these results translate into efficient algorithms for minimum cost admissible queries in the settings of minimum spanning trees, matroids, and matchings in trees; and NP-hardness results in the settings of $s$-$t$ shortest paths and bipartite matching.   
\end{abstract}

\maketitle

%

 
%





\section{Introduction}

We study set selection problems under uncertain weights.
Without uncertainty, an instance of \defi{(min-weight) set selection} consists of a \defi{ground set} $E$, together with weights $w:E\to \RR$, and an (implicitly defined) collection of \defi{feasible sets} $\cF \subseteq 2^E$.
The goal is to find a feasible set $S\in\cF$ which minimizes the total weight $w(S):=\sum_{e\in S} w(e)$.

Set selection encodes many classical problems in combinatorial optimization.
In particular, it encodes the problems of finding minimum spanning trees, shortest paths, and minimum weight perfect matching.  
Maximization problems can be similarly modeled by simply multiplying the weights by $-1$. 

\begin{figure}[ht]
	\begin{subfigure}{.3\textwidth}
		\centering
		\includegraphics[page=4,width=\textwidth]{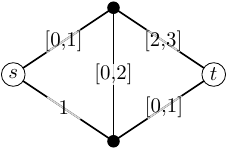}
		\caption{}
		\label{fig:uncertainexample-a}
	  \end{subfigure}%
   \hfill
	  \begin{subfigure}{.3\textwidth}
		\centering
		\includegraphics[page=5,width=\textwidth]{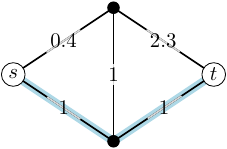}
		\caption{}
		\label{fig:uncertainexample-b}
	  \end{subfigure}%
        \hfill
	  \begin{subfigure}{.3\textwidth}
		\centering
		\includegraphics[page=6,width=\textwidth]{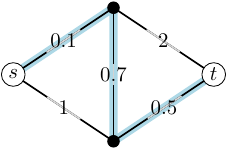}
		\caption{}
		\label{fig:uncertainexample-c}
	  \end{subfigure}
	\centering
\caption{(a) An instance of $s$-$t$ paths with uncertain weights.
(b)+(c) Two realizations of the instance in (a) with different optimal solutions marked in blue.}
	\label{fig:uncertainexample}
\end{figure}

In the uncertain setting, we are again given a ground set $E$ and an implicitly defined collection of feasible sets $\cF\subseteq 2^E$. 
However, we do not have precise weights for each element $e \in E$.
Instead, for each element $e \in E$, we know an interval $I_e := [\ell_e, h_e]$ in which the true weight $w_e$ lies; i.e., $w_e \in I_e$ (see Figure~\ref{fig:uncertainexample-a}).
We call such an interval the \defi{uncertainty interval} of $e$ and denote the collection of intervals by $\cI:= \{ I_e \}_{e \in E}$. 
Note that it is possible for these intervals to be singletons, in that case we call them \defi{trivial}.
Additionally, every weight function $w$ that obeys $\cI$ is called a \defi{realization} of $\cI$; i.e., $w \in \prod_{e \in E}I_e$ (see Figure~\ref{fig:uncertainexample-b} and~\ref{fig:uncertainexample-c}). 
Naturally, realizations of $\cI$ are the possible 
\emph{true weights} of an uncertain instance. 

Under uncertainty, we again are interested in minimizing the weight of the chosen solution.
Note, however, that this could be realization dependent (see Figures~\ref{fig:uncertainexample-b} and~\ref{fig:uncertainexample-c}); i.e., solutions could be optimal for some realizations and not for others.
Thus, the best solution we could hope for is a feasible set that has the best weight independently of the realization.
More formally, we say that a feasible set $S^* \in \cF$ is \defi{universally optimal} if it holds that $w(S^*)\leq w(S)$ for every feasible set $S\in \cF$ and every weight realization $w$ of $\cI$.  

\subsection{Minimum cost queries}
\label{ss:min-queries}

Alas, not every uncertain set selection instance has universally optimal solutions (e.g., Figure~\ref{fig:uncertainexample-a}).
To counteract this, we consider the setting where we can obtain more precise information about the weights at a cost.
The reader can imagine that the original uncertainty intervals represent a rough estimate of the true weights and that we can spend additional resources to find out precise information about them.

The \defi{(non-adaptive) query problem} models exactly the aforementioned situation: 
Combinatorial optimization problems where the weights are uncertain, but can be queried at a cost.
Querying an element reveals its true weight, replacing the possible weights in can take by a unique element.
Here, our objective is to find the least costly set of queries that allows us to compute aa universally optimal solution.

\begin{figure}[ht]
	\begin{subfigure}{.3\textwidth}
		\centering
		\includegraphics[page=1,width=\textwidth]{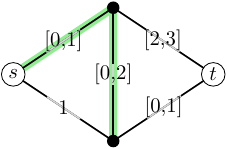}
		\caption{}
		\label{fig:queryexample-a}
	  \end{subfigure}%
   \hfill
	  \begin{subfigure}{.3\textwidth}
		\centering
		\includegraphics[page=2,width=\textwidth]{query-example.pdf}
		\caption{}
		\label{fig:queryexample-b}
	  \end{subfigure}%
        \hfill
	  \begin{subfigure}{.3\textwidth}
		\centering
		\includegraphics[page=3,width=\textwidth]{query-example.pdf}
		\caption{}
		\label{fig:queryexample-c}
	  \end{subfigure}
	\centering
\caption{(a) A minimum-sized admissible query marked in green.
(b)+(c) Different true weights of the queried edges give rise to different universally optimal solutions (marked in blue).}
	\label{fig:queryexample}
\end{figure}

More specifically, we consider the setting where we can make exactly one (non-adaptive) set of queries; i.e., we can select a set $Q\subseteq E$ to be queried, and  for every $e\in E$ we will learn its true weight $w_e\in I_e$.
A set of elements $Q\subseteq E$ is an \defi{admissible query} if after obtaining any possible answer to the queries in $Q$, there exists a universally optimal solution (see Figure~\ref{fig:queryexample-a} for an illustration).%
\footnote{The standard term in the literature seems to be \emph{feasible queries}.
We use the term \emph{admissible queries}, as to make the distinction with the \emph{feasible sets} $S\in \cF$.}
Note that admissible queries always exist, as querying every element is an admissible query.
Querying everything, however, is costly: 
Querying an element $e\in E$ comes at a cost $c_e$, which is known in advance. 
Thus, the objective is to find an admissible query $Q$ that minimizes the total cost $c(Q)$.

We are particularly interested in algorithms that not only compute a minimum cost admissible query, but also compute a universally optimal solution (using the additionally queried information, see Figure~\ref{fig:queryexample-b} and \ref{fig:queryexample-c}).
More concretely, these algorithms operate in the following stages.
\begin{enumerate}
	\item Receive an instance $(E,\cF ,\cI)$.
	\item \emph{Query} a set of elements $Q \subseteq E$.
	\item For the queried elements $e\in Q$, we obtain the true weights $w_e \in I_e$.
	\item Return a universally optimal solution $S\subseteq E$ for the intervals $\cI'$ given by $\{w_e\}_{e \in Q} \cup \{I_e\}_{e \in E\setminus Q}$; i.e., $w(S^*)\leq w(S)$ for all feasible sets $S$ and for all weight realizations $w \in \prod_{e \in Q} \{w_e\} \times \prod_{e \in E\setminus Q} I_e$. 
\end{enumerate}

We highlight that it is \emph{not} required that the algorithm knows the weight of the solution it outputs in the end, only that is \emph{optimal}.
In particular, it is not required that the algorithm has queried all elements in the returned solution.

This model is known as the \defi{explorable uncertainty} model and was introduced by Kahan in 1991~\cite{kahan_model_1991}.
Explorable uncertainty has attracted considerable attention since its introduction; see e.g., the survey of Erlebach and Hoffmann~\cite{erlebach_query-competitive_2015}.
Most of the work has focused on \emph{adaptive} algorithms; i.e., algorithms that can query a few elements, learn their weights, decide whether they query more elements or can compute an optimal solution, and repeat. 
We usually analyze these algorithms via competitive analysis; i.e., they compare against an adversary that knows the true weights beforehand.
Our understanding of the adaptive setting is quite good; almost optimal constant competitive algorithms are known for spanning trees~\cite{erlebach_computing_2008,megow_randomization_2017,merino_minimum_2019}, matroids~\cite{erlebach_query-competitive_2015}, geometric problems~\cite{bruce_efficient_2005}, scheduling~\cite{levi2019scheduling,durr_adversarial_2020,albers2021explorable,dufosse_scheduling_2022,zhang2023scheduling} for slightly different models, and many other settings.
Many of the results can even be obtained in a unified manner by using the technique of \emph{witness sets} as introduced by Bruce et al.~\cite{bruce_efficient_2005}.
Furthermore, for other natural problems such as maximum weight matching, there are impossibility results showing that no constant-competitive algorithms exist~\cite[Chapter 3]{DBLP:phd/basesearch/Meissner18}.  

Surprisingly, we know very little about non-adaptive algorithms. They are used in settings with a huge fixed setup cost for any number of queries.
Merino and Soto~\cite{merino_minimum_2019} studied the problem on matroids with uncertainty sets; i.e., not only intervals.
They provide polynomial time algorithms for finding minimum cost admissible queries.
In particular, for minimum spanning trees, they find minimum cost admissible queries in $\cO(m^2 \alpha(m,n))$ time; here, $n$ is the number of vertices of the graph, $m$ the number of edges, and $\alpha$ is the inverse Ackermann function.
Despite that, their techniques do not seem to easily generalize to other problems.
In particular, they leave open whether one can find minimum cost admissible queries for $s$-$t$ shortest paths or minimum cost perfect matchings in polynomial time.



\subsection{Thresholds of inclusion and exclusion}

\begin{figure}[ht]
	\begin{subfigure}{.3\textwidth}
		\centering
		\includegraphics[page=1,width=\textwidth]{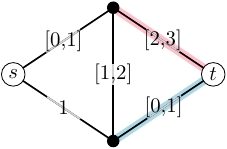}
		\caption{}
		\label{fig:spexample-a}
	  \end{subfigure}%
   \hfill
	  \begin{subfigure}{.3\textwidth}
		\centering
		\includegraphics[page=2,width=\textwidth]{SP_example.pdf}
		\caption{}
		\label{fig:spexample-b}
	  \end{subfigure}%
   \hfill
	  \begin{subfigure}{.3\textwidth}
		\centering
		\includegraphics[page=3,width=\textwidth]{SP_example.pdf}
		\caption{}
		\label{fig:spexample-c}
	  \end{subfigure}
	\centering
\caption{(a) This instance has no universally optimal solution. 
(b) If the edge marked in green has weight less (resp. more) than 2, then it is contained in \emph{every} (resp. no) $s$-$t$ shortest path. 
(c) If the green edge has weight smaller than 1 (resp. larger than $2$), then it is included in all (resp. \emph{no}) shortest $s$-$t$ paths, regardless of the true weights.} 
	\label{fig:spexample}
\end{figure}

To better understand minimum admissible queries, we study a related problem on set selection under uncertainty. 
Note that even if an instance has no universally optimal solution, there are elements which belong to \emph{all} optimal solutions and some others which are in \emph{none} for all realizations (see Figure~\ref{fig:spexample-a}).
Identifying these key elements has proven useful to deal with uncertainty; see e.g.,~\cite{erlebach_computing_2008,megow_randomization_2017,merino_minimum_2019}.

In the setting without uncertainty, these ideas are captured by the concept of \emph{thresholds}.
Here, we consider a fixed element $e \in E$ and are given all weights except the weight of $e$.
Suppose that there are feasible sets with and without $e$: 
When is $e$ contained in an optimal solution?
If the weight of $e$ is very small, say $w_e \approx -\infty$, then $e$ must be included in all optimal solutions.
Similarly, if the weight of $e$ is very large, say $w_e \approx +\infty$, then every optimal solution must avoid $e$. 
The point where this behavior changes is called the \defi{threshold} of $e$, denoted by $T_e$ (see Figure~\ref{fig:spexample-b}).
More formally, $T_e$ is the unique number such that
\begin{itemize}
	\item if $w_e > T_e$, then no optimal solution contains $e$,
	\item if $w_e < T_e$, then $e$ belongs to every optimal solution, 
	\item if $w_e=T_e$, there are optimal solutions with and without $e$.
\end{itemize}
Naturally, the threshold $T_e$ is a function of the weights of all other elements $w_{-e}$.
Furthermore, it is not hard to see that $T_e$ is a piecewise linear function in each coordinate and can be easily computed whenever you can solve the corresponding min-weight set selection problem (see Section~\ref{sec:prelim} for details).

In the uncertain setting, things are a bit more complicated, as we do not have a unique threshold that marks the two regimes (see Figure~\ref{fig:spexample-c}).
For every item $e\in E$, we define the \defi{threshold of inclusion} $\tlow_e$ and the \defi{threshold of exclusion} $\thigh_e$ as follows
\begin{align*}
	\tlow_e& := \inf \{ T_e(w_{-e}) \mid w_{-e} \text{ realization of $\cI \setminus \{I_e\}$}  \} 
 \\
	\thigh_e& := \sup \{ T_e(w_{-e})\mid w_{-e} \text{ realization of $\cI \setminus \{I_e\}$}\}. 
\end{align*}
In other words, $\tlow_e$ is the minimum possible threshold, and $\thigh_e$ is the largest possible threshold.
Note that $\tlow_e = \thigh_e= \infty$ whenever $e$ is in \emph{every} feasible set; similarly, $\tlow_e = \thigh_e= -\infty$ whenever $e$ is in \emph{no} feasible set. 
The interval $[\tlow_e,\thigh_e]$ indicates the possible behavior of $e$ with respect to optimal solutions.
\begin{itemize} 
	\item If $h_{e} < \tlow_e$, then $e$ belongs to any optimal solution for every realization of $\cI \setminus \{I_e\}$. 
	\item If $\ell_{e} > \thigh_e$, then $e$ does not belong to any optimal solution for every realization of $\cI \setminus \{I_e\}$.
	\item Otherwise, there are optimal solutions with and without $e$ for specific realizations of $\cI \setminus \{I_e\}$.
\end{itemize}
Also, the union of $T_e(w_{-e})$ over all realizations $w_{-e}$ of $\cI \setminus \{I_e\}$ is exactly the interval $[\tlow_e, \thigh_e]$. 

The existence of thresholds is central to the isolation lemma, a technique from the 1980s to guarantee uniqueness of optimal solutions; see, e.g.,~\cite{ta-shma_simple_2015} and references therein.
Despite its notoriety, to the best of our knowledge, the problem of computing thresholds under uncertain weights has not been studied prior to this work.

\subsection{Our contribution}

Our main contribution is the introduction of thresholds under uncertainty and using them to analyze minimum cost admissible queries. 
As a first result, we show that computing thresholds and finding minimum cost admissible queries are essentially equivalent.

\begin{itemize}
	\item Given the thresholds of inclusion and exclusion for every element, we can compute a minimum cost admissible query in linear time.
	We can compute a universally optimal solution by solving only one extra set selection problem.
	(See Theorem~\ref{thm:threshold-implies-min-query}.)
	\item Computing thresholds up to a constant additive error reduces to solving a logarithmic number of minimum cost admissible query problems (see Theorem~\ref{thm:min-query-implies-threshold}).
\end{itemize}

After showing this equivalence, we turn towards computing thresholds in various settings.

\begin{itemize}
	\item For the minimum spanning tree problem, we show a $\cO(m \alpha(m,n)+n)$ algorithm for computing all $2m$ thresholds (see Theorem~\ref{thm:thresholds-MST}).
	\item For the maximum weight matching problem on trees, we show a linear time algorithm that given element $e\in E$ it computes its thresholds (see Theorem~\ref{thm:trees}). 
\end{itemize}

In other settings, we are able to show hardness. 
In particular, we show that computing the thresholds is NP-complete for shortest paths on DAGs (see Theorem~\ref{thm:sp}) and for bipartite minimum cost perfect matching (see Theorem~\ref{thm:perfect-matching}). 

As a consequence of the equivalence, we can translate the aforementioned results into the setting of minimum cost admissible queries, obtaining the following results.

\begin{itemize}
	\item An $\cO(m \alpha(m,n)+n)$ time algorithm for finding minimum cost admissible queries in the setting of minimum spanning trees, improving on the $\cO(m^2\alpha(m,n))$ algorithm of Merino and Soto~\cite{merino_minimum_2019}.
	\item A quadratic time algorithm for minimum cost admissible queries for maximum matching on trees. 
	\item Deciding whether a given set is an admissible query for $s$-$t$ shortest paths is NP-complete (see Observation~\ref{obs:hardness-path})
	\item Deciding whether a given set is an admissible query for bipartite minimum cost perfect matching is NP-complete (see Observation~\ref{obs:hardness-matching}).
\end{itemize}

These last two points imply that there is no polynomial time algorithms for finding minimum cost admissible queries in the setting of shortest paths and bipartite minimum cost perfect matching (unless $\text{P}=\text{NP}$).

\subsection{Further related work}
\label{ss:related-work}

Besides what was mentioned in Subsection~\ref{ss:min-queries}, there are a couple of other explorable uncertainty settings related to this work.

\textbf{Stochastic uncertainty.}
In this model instead of working only with uncertainty intervals as prior information on the weights, they are given additional stochastic information.
Again, the objective is to minimize the cost of querying while being able to compute an optimal solution. 
Particularly relevant to us is the work of Megow and Schlöter that explores adaptive algorithms for set selection problems~\cite{megow_set_2023}.

\textbf{Value-based models.}
There are models with the additional requirement that the produced solution must belong to the set of queried element; see e.g., the work of Singla~\cite{singla_price_2018}.
In other words, the algorithm not only needs to produce an optimal solution, but also the value of the solution.

The optimality condition is sometimes relaxed, allowing to output a solution that is additively suboptimal by some pre-specified constant; see e.g., the work of Feder et. al.~\cite{feder_computing_2000,feder_computing_2007}.
We highlight that these last works are non-adaptive, however since their setting requires finding the value, it has a very different flavor.



\section{Threshold preliminaries}
\label{sec:prelim}

We begin by analyzing basic properties of the threshold functions in both without and with uncertainty.

\subsection{Thresholds without uncertainty}

\begin{figure}[ht]
	\begin{subfigure}{.5\textwidth}
		\centering
		\includegraphics[page=1,width=\textwidth]{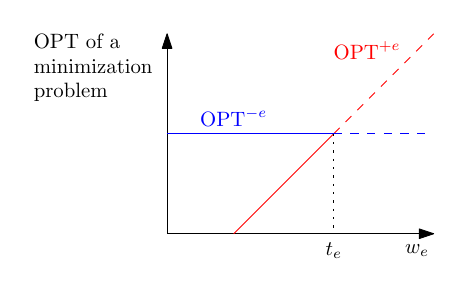}
		\caption{}
		\label{fig:threshold-a}
	  \end{subfigure}%
	  \begin{subfigure}{.5\textwidth}
		\centering
		\includegraphics[page=2,width=\textwidth]{graphs.pdf}
		\caption{}
		\label{fig:threshold-b}
	  \end{subfigure}
	\centering
\caption{(a) Behaviour of $\OPT^{-e}$ and $\OPT^{+e}$ when we change the weight of $e$.
The threshold $T_e$ is exactly the point at which these two values are the same.
(b) Schematic view of this threshold as a function of the weight of some other element $f$.}
	\label{fig:threshold}
\end{figure}

Note that for every $e \in E$ can easily compute $T_e$  as follows.
Let $\textrm{OPT}^{-e}(w_{-e})$ be the weight of an optimal feasible set $S$ under the constraint $e\not\in S$, and let $\textrm{OPT}^{+e}(w_{-e})$ be the similar weight under the constraint $e\in S$ and imposing $w_e=0$.
(See Figure~\ref{fig:threshold-a}.) We have
\[
	T_e(w_{-e}) = \textrm{OPT}^{-e}(w_{-e}) - \textrm{OPT}^{+e}(w_{-e}).
	\]

How does each of the terms $\OPT^{-e}(w_{-e})$ and $\OPT^{+e}(w_{-e})$ depend on the weight of another element $f \in E$? 
Consider the optimal feasible set without $e,f$, and the optimal feasible set without $e$ but with $f$. 
The weight of the first is constant in $w_f$, while the second is linear (with slope 1) in $w_f$.
Thus, $\textrm{OPT}^{-e}(w_{-e})$ is a constant function of $w_f$ up to some point, from which it has a unit slope. 
Similarly, $\textrm{OPT}^{+e}(w_{-e})$ is a similar function in $w_f$, but with a different constant part and different threshold. 
Hence, their difference $T_e(w_{-e})$ is a piecewise linear function in $w_f$, consisting of three parts. 
The first and third parts are constant, while the second part has a slope of either $+1$ or $-1$, depending on how the thresholds compare to each other (see Figure~\ref{fig:threshold-b}). 
As a consequence, we have that $T_e$ is continuous and monotone in each coordinate.

The next lemma states that $\OPT^{+e}$ and $\OPT^{-e}$ are structurally very similar. 
Intuitively, this holds as we are optimizing a linear problem along the same direction and only making a small change in the restrictions.

We formalize this in the context of polytopes. 
Given a set of vectors $X\subseteq \RR^d$ its \defi{convex hull} $\conv(X)$ consists of all convex combinations of vectors in $X$.
For $F \subseteq E$ we denote by $\chi_F\in\{0,1\}^E$ the \defi{characteristic vector} of $F$.
For $x,y \in \{0,1\}^n$ their \defi{Hamming distance} is $d(x,y) = \sum_{i=1}^n |x_i-y_i|$.
\begin{lemma}
	\label{lem:polytope}
	
	Let $P:= \conv(\{\chi_F \mid F \in \cF \})$, $w \in \RR^E$, and $e\in E$.
	There exists $x',y' \in \{0,1\}^{E}$ such that $w\cdot x' = \min \{ w\cdot x \mid x\in P, \; x_e = 0 \}$, $w \cdot y' = \min \{ w\cdot x \mid x\in P, \; x_e = 1 \}$, and $xy$ is an edge of the polytope $P$.
\end{lemma}

\begin{proof}
	Choose two vertices $x', y' \in \{0,1\}^E$ of $P$ such that $x'\in \argmin \{w\cdot x \mid x \in P,\; x_e =0  \}$, $y'\in \argmin \{w\cdot x \mid x \in P,\; x_e =1  \} $, and the Hamming distance between $x'$ and $y'$ is minimized.
	Suppose that $x'$ and $y'$ are not adjacent in $P$. 
	Then, there exists $\mu \in (0,1)$ such that $z:=\mu x' + (1-\mu)y'$ is a convex combination of the vertices of $P$ except $x'$ and $y'$; i.e., $z = \sum_{i=1}^k\lambda_iz^i$ where for every $i\in[k]$ we have $\lambda_i \in (0,1)$ and that $z^i \in \{0,1\}^E$ is a vertex of $P$ different from $x'$ or $y'$.

	Let us define $I_0 = \{i \in [k] \mid z^i_e = 0\}$ and $I_1 = \{i \in [k] \mid z^i_e = 1\}$.
	Note that $\sum_{i \in I_0} \lambda_i = \mu$ and $\sum_{i \in I_1} \lambda_i = (1-\mu)$.
	Furthermore, for every $i \in I_0$ we have that $w\cdot z^i = w\cdot x'$ and for every $i \in I_1$ we have that $w\cdot z^i = w\cdot y'$, as otherwise we have that
\[
	w\cdot z = w \cdot \left(\sum_{i\in I_0}\lambda_i z^i + \sum_{i \in I_1}\lambda_i z^i \right) > \mu(w\cdot x)+(1-\mu)(w\cdot y) = w\cdot z,
\] 
which is not possible.
Thus, we conclude that $z^i \in \argmin \{w\cdot x \mid x \in P, x_e =1 \}$ for $i \in I_1$ and that $z^i \in \argmin \{w\cdot x \mid x \in P, x_e =0 \}$ for $i \in I_0$.

Let us define $A$ as the coordinates in which $x'$ and $y'$ agree; i.e., $A= \{f\in E \mid x'_f=y'_f\}$.
Note that for $f \in A$ we have that $z^i_f = x'_f = y'_f$ for every $i \in [k]$.
Furthermore, $z^i = x'$ whenever $i \in I_0$;
otherwise, there exists $i^* \in I_0$ such that $z^{i^*}$ agrees in more than $|A|$ coordinates with $y'$, hence  $d(z^{i^*},y') < |E|-|A|  = d(x',y')$ and contradicting the minimality of $d(x',y')$. 
A similar argument shows that $z^i = y'$ whenever $i \in I_1$.
This contradicts the fact that $z^i$ was not $x'$ nor $y'$ for every $i \in [k]$.
\end{proof}

Lemma~\ref{lem:polytope} has interesting consequences when looking at specific problems. 
In these problems edges of the corresponding polytope are well-understood.
For example, for minimum spanning trees, a result due to Hausmann and Korte implies that two trees are adjacent if and only if they differ in the exchange of two edges. Combining this with Lemma~\ref{lem:polytope} we obtain that.
 
\begin{corollary}
    \label{cor:close-mst}
    Let $G = (V,E)$ be a graph with weights. 
    There exists an edge $f \in E-e$ and a spanning tree $T^+$ such that $T^+$ is a spanning tree that contains $e \in E$ of minimum weight, and $T^-:= T^+-e+f$ is a spanning tree that avoids $e \in E$ of minimum weight.
\end{corollary}

Analogous statements also hold for other set selection problems; see e.g., matchings and perfect matchings~\cite{MR371732}, $s$-$t$ paths and flows~\cite{MR555751}.

\subsection{Thresholds with uncertainty}

A weight realization $w$ such that $w_f\in\{\ell_{f}, h_{f}\}$ for every $f\in E$ is said to be \defi{extreme}.
Observe that by monotonicity of $T_e$ --- which can be non-increasing or non-decreasing, see Figure~\ref{fig:threshold} --- we have $\tlow_e=T_e(w_{-e})$ for some extreme realization $w_{-e}$.
The same holds for $\thigh_e$.

This means that the threshold of inclusion can be computed by minimizing over $2^{|E|}$ different extreme weights, which implies the following observation.

\begin{observation}	\label{obs:in-NP}
	Suppose that the underlying optimization problem can be solved in polynomial time for a fixed weight realization. 
	The decision version of computing the threshold of inclusion asks if $\tlow_e$ is at most some given value.
	The decision version of computing the threshold of exclusion asks if $\thigh_e$ is at least some given value. 
	Both problems are in NP, and the certificates are the corresponding extremal realizations.
\end{observation}

We can obtain an alternative characterization of the threshold of inclusion. For a given realization $w_{-e}$, let $S^-\in \cF$ be the feasible solution with optimal value $\textrm{OPT}^{-e}(w_{-e})$, and $S^+\in \cF$ be the feasible solution with optimal value $\textrm{OPT}^{+e}(w_{-e})$. Then we can assume without loss of generality that the realization sets $w_f$ to its lower bound for all $f\in S^-$, and to its upper bound for all $f\not\in S^-$. In other words $\tlow_e$ is the minimum among all $S\in \cF$ with $e\not\in S$ of $\ell(S) - \textrm{OPT}^{+e}(w)$, where $w_e=0$, and $w_f=\ell_f$ for all $f\in S$ and $w_f=h_f$ otherwise, conditioned on $S$ being an optimal solution with respect to these weights $w$ and the constraint $e\not\in S$.

Similarly, the threshold of exclusion can be expressed as the maximum over all $T\in \cF$ with $e\in T$ of the difference $ \textrm{OPT}^{-e}(w) - \ell(T)$, where  $w_f=\ell_f$ for all $f\in T$ and $w_f=h_f$ otherwise, conditioned on $T$ being an optimal solution with respect to these weights $w$ and the constraint $e\in T$.

\section{From thresholds to minimizing queries, and back}
\label{sec:min_query}

In this section, we prove that the problem of computing thresholds and finding minimum cost admissible queries are effectively equivalent. 

\subsection{The query problem reduces to threshold computation}

We begin by classifying elements according to their relation to thresholds.
We say that $e \in E$ is \defi{blue} if $I_e$ is non-trivial and $h_e \leq \tlow_e$. 
Similarly, we say that $e\in E$ is \defi{red} whenever $I_e$ is non-trivial and $\thigh_e \leq \ell_e$.
Intuitively, blue elements are \emph{safe} to include in an optimal solution, and red elements are \emph{safe} to delete from optimal solutions.

We collect some simple observations.

\begin{lemma}
	No element $e \in E$ is blue and red.
\end{lemma}
\begin{proof}
	Let $e \in E$ be an element that is blue and red.
	Thus, $h_e \leq \tlow_e \leq \thigh_e \leq \ell_e $, 	which implies that $I_e$ is trivial, a contradiction.
\end{proof}
Therefore, the elements are partitioned into blue, red, trivial, and uncolored non-trivial.
We also note that colors are preserved under queries.
\begin{lemma}
    Let $Q\subseteq E$ be a set to be queried. 
    If $e \notin Q$ was blue (resp. red) before querying $Q$, then it is blue (resp. red) after querying $Q$.
\end{lemma}
\begin{proof}
	We consider an instance with uncertainty intervals $\cI = \{I_e\}_{e \in E}$.
	After querying a set $F\subseteq E$ and obtaining the true weights $w_e \in I_e$ for $e\in E$, we have a new instance with uncertainty intervals $\cI' : = \{ \{w_e\} \}_{e \in F} \cup \{I_e\}_{e \in E\setminus F}$.
	For every $e\in E$, let $\tlow_e,\thigh_e$ (resp. $\tlow_{e,Q},\thigh _{e,Q}$) be the thresholds for $e$ with intervals $\cI$ (resp. $\cI'$).
	Note that every realization of $\cI'$ is a realization of $\cI$.
	Thus, $\tlow_e \leq \tlow_{e,Q}$ and $\thigh_{e,Q} \leq \thigh_{e}$ and the lemma follows.
\end{proof}

Furthermore, as already noted by Merino and Soto~\cite{merino_minimum_2019}, in the context of matroids, supersets of admissible queries are admissible.

\begin{lemma}
	\label{lem:supersets}
	If $F\subseteq E$ is an admissible query, all its supersets $F \subseteq F'$ are admissible queries. 
\end{lemma}

The key observation is that uncolored non-trivial elements \emph{must} be queried.
\begin{lemma}
	\label{lem:must-query}
	If $F \subseteq E$ is an admissible query, then it contains the set
	\[Q = \{ e\in E \mid e \text{ is uncolored and $I_e$ is non-trivial} \}.\]
\end{lemma}
\begin{proof}
	Suppose, aiming at a contradiction, that there is an admissible query $F$ and an element $e^* \in Q\setminus F$. Let $A_{e^*}=(\ell_{e^*}, h_{e^*})\cap [\tlow_{e^*},\thigh_{e^*}]$. Since $h_{e^*}>\tlow_{e^*}$ and $\ell_{e^*}<\thigh_{e^*}$, the set $A_{e^*}$ is nonempty. Choose $w^*\in A_{e^*}$  and  and $\epsilon > 0$ such that $(w^*-\epsilon,w^*+\epsilon) \subseteq I_{e^*}$. 
	Since $\tlow_{e^*} \le w^* \le \thigh_{e^*} $, we have that there is a realization $w \in \RR^E$ such that $e^*$ is in an optimal solution $S^+$, $e^*$ avoids an optimal solution $S^-$, and $w_{e^*} = w^*$.
	Furthermore, we define $w^- \in \prod_{e \in E} I_e$ as $w$ except that in the coordinate $e^*$ we have that $w^-_{e^*} = w^*+\epsilon$. 
	Similarly,  we define $w^+ \in \prod_{e \in E} I_e$ as $w$ except that in the coordinate $e^*$ we have that $w^+_{e^*} = w^*-\epsilon$.  
	
	Suppose now we query $E-e^*$. 
	Since $F\subseteq E-e^*$ was an admissible query, by Lemma~\ref{lem:supersets} we obtain that $E-e^*$ is also admissible.
	Thus, even if we reveal the weights on $E-e^*$ to be $w_{-e}$, there exists a feasible set $S \in \cF$ that is optimal independently of the value of $e^*$.
	If the true value of $e^*$ is $w^*$, we have that $w(S)= w(S^-) = w(S^+)$, as they are all optimal solutions.
	Now, if the true value of $e^*$ is $w^*-\epsilon$, we have that $e^* \in S$ (otherwise $w^+(S^+)<w^+(S)$). 
	Finally, if the true value of $e^*$ is $w^*+\epsilon$, we have that $e^* \not\in S$ (otherwise $w^-(S^-)<w^-(S)$). 
	Thus, we have that $e^* \in S$ and $e^*\notin S$, a contradiction.
\end{proof}

In view of Lemma~\ref{lem:must-query}, we only need to figure out what to do with blue and red elements.
The next lemma shows that it is always safe to pick blue elements in optimal solutions and avoid red elements.
\begin{lemma}
	\label{lem:contain-blue-avoid-red}
	For every realization $w \in \prod_{e \in E} I_e$, there is a $w$-optimal solution $S^*$ such that 
	\[\{e \in E \mid \text{$e$ is blue} \} \subseteq S^* \subseteq \{e \in E \mid \text{$e$ is not red}\} .\]
\end{lemma}
\begin{proof}
	Let $B : = \{e \in E \mid  \text{$e$ is blue}\}$ and $R:= \{e \in E \mid \text{$e$ is red}\}$.
	Let $S$ be a $w$-optimal solution. We also define 
	\[g := \min \{w(S') \mid S'\subseteq E \text{ and } B\subseteq S' \subseteq E \setminus R  \} -w(S)  \geq 0 . \]
	Assume, for the sake of contradiction, that $g > 0$.

	Let $\epsilon := \frac{g}{4|E|}$.
	For $e \in B$, choose $\delta_e \in (-\epsilon ,\epsilon )$ such that $w_e + \delta_e < \tlow_e$ and $w_e + \delta_e \in I_e$.
	Similarly, for $e \in R$, we choose $\delta_e \in (-\epsilon ,\epsilon )$ such that $w_e + \delta_e > \thigh_e$ and $w_e + \delta_e \in I_e$.
	For the remaining elements $e \in E\setminus (B \cup R)$, we set $\delta_e = 0 $.

	We now define $\hat w \in \prod_{e \in E} I_e$ as $\hat w := w+\delta$.
	Let $\hat S$ be a $\hat w$-optimal solution. 
	Since for $e \in B$ we have that $\hat w_e < \tlow_e$, this implies that $B\subseteq \hat S$.
	Similarly, for $e \in R$ we have that $\hat w_e > \thigh_e$, which implies that $\hat S \subseteq E \setminus R$.
	Therefore,
	\begin{equation}
		\label{eq:lower-bound-g}
		w(\hat S)-w(S)\geq \min \{w(S) \mid S\subseteq E \text{ and } B\subseteq S \subseteq E \setminus R  \} - w(S) = g.
	\end{equation}
	Since $\hat S$ is $\hat w$-optimal, we have that $\hat w(\hat S) \leq \hat w(S)$.
	As a consequence, we have that $w(\hat S) + \delta(\hat S) \leq w(S) + \delta(S)$.
	Thus,
	\begin{equation}
		\label{eq:upper-bound-g-half}
		w(\hat S) -w(S)  \leq \delta(S)-\delta(\hat S) \leq \epsilon (|S|+|\hat S|) \leq 2\epsilon |E| = \frac{g}{2}.
	\end{equation}
	Combining \eqref{eq:lower-bound-g} with \eqref{eq:upper-bound-g-half} we obtain that $\frac{g}{2} \geq g$, a contradiction.
\end{proof}

We can combine Lemma~\ref{lem:must-query} and Lemma~\ref{lem:contain-blue-avoid-red} to show that the uncolored non-trivial elements are the \emph{unique} minimum-sized admissible query.
Therefore, there is a natural algorithm for finding minimum cost admissible queries: Query the uncolored non-trivial elements and every element of negative cost.
We can extend this to an algorithm that also computes universally optimal solutions as follows.

\begin{algo}{Minimum cost admissible queries and universally optimal solutions}
	\label{alg:min-cost-query}
	\begin{enumerate}[leftmargin=8mm, noitemsep, topsep=3pt plus 3pt]
		\item For every $e\in E$, compute the thresholds $\tlow_e , \thigh_e$. 
		\item Compute $B \gets \{ e \in E \mid \text{$e$ is blue}\}$, $R \gets \{ e \in E \mid \text{$e$ is blue}\}$, 
		
		$T \gets \{e \in E \mid \text{$I_e$ is trivial}\}$ and $Q \gets E\setminus (B\cup R \cup T)$. 
		\item \emph{Query} $Q$ and every element of negative cost.

		Obtain true weights $w_e \in I_e$ for $e \in Q$.
		\item Let $M \gets \sum_{e \in T \cup Q}|w_e|+\sum_{e \in B \cup R} |\ell_e|+|h_e|$.
		\item For every $e \in B$ assign $w_e \gets -M$ and for every $e \in R$ assign $w_e \gets M$. 
		
		\item \emph{Return} an element of $\argmin \{ w(S) \mid S \in \cF \}$ as a universally optimal solution. 
	\end{enumerate}
\end{algo}
	
	\begin{theorem}
	\label{thm:threshold-implies-min-query}
	Suppose all thresholds for $(E,\cF,\cI)$ can be computed in time $T_{THR}$ and that the optimization problem on $(E,\cF)$ can be solved in time $T_{OPT}$.
	We can find a minimum cost admissible query problem in time $\cO(T_{THR}+|E|)$, and after the weights of the queries are revealed we can find a universally optimal solution in time $\cO(T_{\OPT}+|E|)$
\end{theorem}
\begin{proof}
	We show that Algorithm~\ref{alg:min-cost-query} solves the minimum cost admissible query problem, since 
	it is clear that Algorithm~\ref{alg:min-cost-query} runs in the desired running time. 

	We now show that $Q$ is a admissible query. 
	After Algorithm~\ref{alg:min-cost-query} queries $Q$, it obtains true values $w_e \in I_e$.
	We aim to show that the algorithm returns an optimal solution for every realization $w$ of the intervals after querying $\cI'$; that is $\cI' := \{I_e\}_{e \in E\setminus Q} \cup \{\{w_e\}\}_{e\in Q}$.
	Let $M := \sum_{e \in T \cup Q}|w_e|+\sum_{e \in B \cup R} |\ell_e|+|h_e|$.
	We also define $w_e = -M$ for blue $e$ and $w_e = M$ for red $e$ as in Algorithm~\ref{alg:min-cost-query}.
	Let $S'$ be a $w$-optimal solution.
	It is easy to see that $S'$ contains every blue element, and it avoids every red element.
	Choose any $w^* \in \cI'$.
	By Lemma~\ref{lem:contain-blue-avoid-red}, we have that there is a $w^*$-optimal solution $S^*$ such that 
	\[\{e \in E \mid \text{$e$ is blue} \} \subseteq S^* \subseteq \{e \in E \mid \text{$e$ is not red}\} .\]
	Note that since $w|_Q=w^*|_Q$ and $w|_T = w^*|_T$, we have that
	\[-M|B| + w(S^* \cap (Q\cup T)) = w(S^*) \geq w(S') = -M|B|+w(S' \cap (Q\cup T)) .\]
	We conclude that $w^*(S' \cap (Q\cup T)) \leq w^*(S^* \cap (Q\cup T))$.
	Using that we obtain
	\[w^*(S') = w^*(B)+ w^*(S' \cap (Q\cup T)) \leq w^*(B) + w(S^* \cap (Q\cup T)) = w^*(S^*). \]
	Hence, $S'$ is $w^*$-optimal, and in turn $Q$ is an admissible query.

	The fact that $Q$ is of minimum cost follows directly from Lemma~\ref{lem:must-query}.
\end{proof}
\subsection{Threshold computation reduces to the query problem.}

We show now how to additively approximate thresholds by using minimum cost admissible queries. 
In particular, we only use a weaker oracle that tests whether $E-e$ is an admissible query.
Naturally, this oracle can be simulated with a minimum cost admissible query oracle by using the costs $c_e = 1$ and $c_f = 0$ for $f \in E-e$.

We reason for $\tlow_e$ in this section, as the arguments for computing $\thigh_e$ are dual.
The main idea is that for a given $e\in E$ and  $\alpha \in \RR$ we can check whether $\tlow_e < \alpha$ by using the aforementioned oracle.
To this end, given uncertainty intervals $I_f$ for $f\in E-e$, we define $K = -\sum_{f \in E-e} |\ell_f|+|h_f|$, $I^\alpha_e = [K-1,\alpha]$ and $\cI^\alpha = \{I_f\}_{f \in E-e} \cup \{I_e^\alpha\}$.
Note that $\tlow_e$ is the same for $\cI$ and $\cI^\alpha$. 
First, we show that we can assume that $\thigh \neq -\infty$.

Recall that we can test whether $\thigh_e \neq -\infty$ by running one optimization problem and setting the weight of $e$ to be very small, namely the negation of the total weight upper bounds.
Thus, we assume in the next lemma that $\thigh_e \neq -\infty$

\begin{lemma}
	\label{lem:binary-search}
	$\tlow_e \geq \alpha$ for $\cI$ if and only if $E-e$ is an admissible query for $\cI^\alpha$.
\end{lemma}
\begin{proof}
	If $\tlow_e \geq \alpha = h_e$, then $e$ is blue, and therefore $E-e$ is an admissible query.

	For the converse, note that $\tlow_e < \alpha = h_e$, implies that $e$ is not blue.
	Furthermore, since $\thigh_e \neq -\infty$, then $\thigh_e \geq \ell(E-e)$, implies that $e$ is not red.
	Hence, $e$ is uncolored and must be queried, which implies that $E-e$ is not admissible.
\end{proof}

As a consequence, we can simply apply Lemma~\ref{lem:binary-search} and obtain an algorithm that computes thresholds with additive error.

\begin{theorem}
	\label{thm:min-query-implies-threshold}
	Suppose we can decide whether $E-e$ is an admissible query in time $T$, and that we can solve the corresponding optimization problem in time $T_{OPT}$. 
	For every $e \in E$ and every additive precision $\delta> 0 $, we can compute $a,b \in [-K,K]$ such that $\tlow_e \in [a,a+\delta]$ and $\thigh_e \in [b,b+\delta]$ (or decide whether $T^-_{e},T^+{e} = \pm\infty$).
	This algorithm runs in time $\cO\left(T \log(\frac{2K}{\delta})+T_{OPT}\right)$.
\end{theorem}
\begin{proof}
	We only need to check beforehand whether the thresholds are $\pm \infty$ and then run the corresponding binary search.
\end{proof}

\section{Minimum spanning trees}




In this section we consider the classical minimum spanning tree problem (MST). 
A \defi{minimum spanning tree} of a graph $G = (V,E)$ with weights $w : E\to\RR$ is a spanning tree $T$ of $G$ that minimizes $w(T)$.
We show algorithms for computing the thresholds of inclusion and exclusion for MSTs.
We assume that the edge $e = uv$ for which we want to compute the weights is not a \emph{bridge}, also known as \emph{cut-edge}, otherwise both thresholds are $+\infty$.

\subsection{Thresholds via greedy}

We begin by giving a simple greedy-based algorithm for computing thresholds.

First, we recall the classical \defi{red} and \defi{blue} rules for computing MSTs. 
\begin{itemize}
	\item Red rule: edge $e$ is in no MST if and only if $w_e > \min\limits_{C\text{ cycle},\; e \in C} \max\limits_{f \in C-e} w_f$.
	\item Blue rule: edge $e$ belongs to all MST if and only if $w_e<\max\limits_{C\text{ cut},\;e\in C} \min\limits_{f\in C-e} w_f$.
\end{itemize}
Direct application of the red rule implies that \(\tlow_e\) is the minimum over all cycles $C$ containing $e$ of \(\max \ell_{f}\) over \(f\in C - e\).
Similarly, by applying the blue rule, we obtain \(\thigh_e\) is the maximum over all cuts $C$ containing $e$ of \(\min h_{f}\) over \(f\in C-e\). As a result one can compute  $\tlow_e$ and $\thigh_e$ via Kruskal's algorithm and reverse-Kruskal.

The standard implementation of these procedures achieves a running time of $\cO(m\log n)$ and $\cO(m \log n (\log\log n)^3 )$ for computing $\thigh_e$ and $\tlow_e$ respectively~\cite{thorup_near-optimal_2000}.

We note that the ideas in this subsection generalize straightforwardly to the \emph{matroid} setting, obtaining efficient algorithms for computing the thresholds of inclusion or inclusion. 

\subsection{Thresholds via bottleneck paths}

Given an undirected graph $G=(V,E)$ with weights $w \in \RR^E$ and a path $P$, the \defi{bottleneck of the path} $P$ is given by $\max_{e \in P} w_e$.
A \defi{bottleneck path} $P$ between $u,v \in V$ is a path of minimum bottleneck.
The \defi{bottleneck between $u$ and $v$} is the bottleneck of a bottleneck $u$-$v$ path.
A classical result of Hu~\cite{93048544-4891-3daf-a365-99ecc24ba101} states that the paths in a minimum spanning tree are bottleneck paths.

\begin{theorem}[\cite{93048544-4891-3daf-a365-99ecc24ba101}]
	\label{thm:bottleneck-MST}
	Let $G=(V,E)$ be a graph with weights $w \in \RR^E$ and let $T$ be an MST of $G$.
	For every $u,v \in V$ the unique $u$-$v$ path in $T$ is a $u$-$v$ bottleneck path. 
\end{theorem}

We consider a slight variation of the bottleneck problem.
We say that a path is \defi{trivial} if it is of length one, otherwise, we say it is \defi{non-trivial}.
The non-trivial bottleneck between $u$ and $v$, denoted by $b_w(u,v)$, is the minimum bottleneck of a non-trivial $u$-$v$ path; i.e., 
\[ b_w(u,v)  = \min_{\text{$P$ $u$-$v$ nontrivial path}} \max_{e \in P} w_e .\]
In other words, the non-trivial bottleneck between $u$ and $v$ in $G$, is the bottleneck between $u$ and $v$ in $G-uv$.

We show that computing non-trivial bottlenecks is equivalent to computing thresholds.
\begin{lemma}
	\label{lem:bottleneck-threshold}
	Let $G=(V,E)$ be a graph with weights $w$.
	For every $uv \in E$, we have that $T_{uv}(w_{-uv})=b_{w}(u,v)$.
\end{lemma}
\begin{proof}
	Let $T$ be a minimum spanning tree, we consider two cases.
\begin{itemize} 
	\item \textbf{Case 1:} $e\notin T$. 
	Let $T'$ be $\OPT^{+e}(w_{-e})$.
	By Corollary~\ref{cor:close-mst} we may assume that $T'=T+e-f$.
	Let $P$ be the unique $u$-$v$ path in $T$.
	Furthermore, $f$ must be an edge of maximum weight in $P$.
	Otherwise, there is an edge $f' \in P$ such that $w_{f'}>w_{f}$, which implies that $w(T+e-f')<w(T+e-f)$, a contradiction.
	Since $P$ was a $u$-$v$ bottleneck path by Theorem~\ref{thm:bottleneck-MST}, we conclude that 
	\[ T_e(w_{-e})=\OPT^{-e}(w_{-e})-\OPT^{+e}(w_{-e})=w_f= b_w(u,v) .\]

\item	\textbf{Case 2:} $e\in T$. 
	Let $T'$ be $\OPT^{-e}(w_{-e})$.
	By Corollary~\ref{cor:close-mst} we may assume that $T'=T-e+f$.
	Note that $T'$ is an MST of $G-e$.
	Let $P$ be the unique $u$-$v$ path in $T'$.
	As before, $f$ must be an edge of maximum weight in $P$.
	Otherwise, there is an edge $f' \in P$ such that $w_{f'}>w_{f}$, which implies that $w(T'+e-f')<w(T'+e-f)=w(T)$, a contradiction.
	Since $P$ was a $u$-$v$ bottleneck path in $G-e$ by Theorem~\ref{thm:bottleneck-MST}, then it is a non-trivial $u$-$v$ path of minimum bottleneck in $G$.
 	Thus, we conclude that 
	\[ T_e(w_{-e})=\OPT^{-e}(w_{-e})-\OPT^{+e}(w_{-e})=w_f= b_w(u,v).\qedhere\]
	\qedhere\end{itemize}
\end{proof}

Combining Lemma~\ref{lem:bottleneck-threshold} with the fact that $b_{w}(u,v)$ is increasing with respect to $w$, we obtain that for every $uv \in E$ it holds that $\tlow_{uv} = b_{\ell}(u,v)$
and $\thigh_{uv} = b_{h}(u,v)$. 
Thus, the problem of computing the threshold for $e \in E$ is equivalent to computing $b_\ell(u,v)$ and $b_h(u,v)$. 

We now provide a fast algorithm for computing all thresholds in the minimum spanning tree setting.
Given a minimum spanning tree $T$ and $e \in T$, we say that $f$ is a \defi{replacement} of $e$ in $T$ if $T+f-e$ is a minimum spanning tree of $G-e$.
Furthermore, if the replacement of $uv$ is $xy$ and $x$ is connected to $u$ in $T-e$, we say that $x$ is the replacement of $u$.
Replacements can be computed very efficiently by using the technique of path compression, as shown by Tarjan~\cite{MR0545544}.

\begin{lemma}
	Given a weighted graph and a minimum spanning tree $T$, we can compute all $n-1$ replacements of $T$ in time $\cO(m\alpha(m,n)+n)$.
\end{lemma}

We also make use of the computation of bottlenecks in a tree. 
This is usually a subroutine of minimum spanning tree verification algorithms, originally shown to run in linear time by Dixon, Rauch, and Tarjan~\cite{MR1192301}, with further simplifications by King~\cite{MR1441172} and Hagerup~\cite{MR2587710}.

\begin{lemma}
	Given a tree $T$ with weights on the edges and $k$ pairs of vertices $(u_1,v_1),\dots, (u_k,v_k)$, we can compute the bottlenecks in $T$ of the $k$ pairs $b_T(u_1,v_1),\dots, b_T(u_k,v_k)$ in time $\cO(n+k)$.
\end{lemma}

The last ingredient we need is Chazelle's algorithm for minimum spanning trees~\cite{chazelle_minimum_2000}.
\begin{lemma}
	Given a weighted graph, its minimum spanning tree can be computed in time $\cO(m\alpha(m,n))$.
\end{lemma}

We combine the aforementioned algorithms to obtain the following algorithm for non-trivial bottlenecks.
\begin{algo}{Non-trivial bottlenecks}
\label{alg:mst-threshold}
\begin{enumerate}
	\item Compute a minimum spanning tree $T$ via Chazelle's algorithm.
	\item For every edge $e\in T$, compute its replacement $r_T(e)$ via path compression. 
	\item Using the tree-bottleneck algorithm, we compute $b_T(u,v)$ for $(u,v) \in E\setminus T$, and for every $uv \in T$ we compute $b_T(x,u)$ where $x$ is the replacement of $u$ and $y$ the replacement of $v$.
	\item For $uv \notin T$, return $b_T(u,v)$. Otherwise, let $x$ be the replacement of $u$ and $y$ the replacement of $v$, and return $\max \{b_T(u,x), w(xy), b_T(v,y) \}$.
\end{enumerate}
\end{algo}

We can use Algorithm~\ref{alg:mst-threshold} using $\ell$ (resp. $h$) as weights to compute all thresholds $\tlow$ (resp. $\thigh$) for every $e \in E.$

\begin{theorem}
	\label{thm:thresholds-MST}
	For the minimum spanning tree problem, we can compute all thresholds of inclusion and exclusion in time $\cO(m\alpha(m,n)+n)$.
\end{theorem}
\begin{proof}
	By Lemma~\ref{lem:bottleneck-threshold} it suffices to compute non-trivial bottlenecks for every $e \in E$ with weights $h$ and $\ell$.
	It is clear that Algorithm~\ref{alg:mst-threshold} runs in the desired time, thus we only prove correctness.

	Let $T$ be the minimum spanning tree used by Algorithm~\ref{alg:mst-threshold}.
	For $uv \notin T$, it is clear that $T$ is a minimum spanning tree of $G-uv$.
	Thus, $b_T(u,v)$ is the non-trivial bottleneck for $uv$ by Theorem~\ref{thm:bottleneck-MST}.
	If $uv \in T$, let $x$ be the replacement of $u$ in $T$ and $y$ the replacement of $v$ in $T$.
	Let $P^{ux}$ be the unique $ux$-path in $T$, and $P^{yv}$ be the unique $yv$-path in $T$.  
	Then, $T':=T+xy-uv$ is a minimum spanning tree of $G-uv$ and the unique $uv$ path in $T'$ is $P':=P^{ux}P^{yv}$. 
	Since the bottleneck of $P'$ is $\max \{b_T(u,x),w(xy),b_T(y,v)\}$ we conclude by Theorem~\ref{thm:bottleneck-MST}. 
\end{proof}
\section{Shortest paths}

We consider the problem of computing a shortest path between two given vertices $s,t$, with non-negative uncertain edge weights.
In~\cite{feder_computing_2007} this problem has been studied in a model, where the algorithm needs to compute the length of the shortest path using as few queries as possible. 
However, in our setting it is enough to produce some path guaranteed to be the shortest one, even though its exact length could be uncertain.

Consider an edge $e=(u,v)$.
The threshold of exclusion $\thigh_e$ can be stated as the maximum over all weight realizations $w$ of the difference between the length $\ell$ of a shortest path from $s$ to $t$ without using $e$ and the length of a shortest path $P$ from $s$ to $t$ forced to use $e$. 

Without loss of generality we can assume that all edges of $P$ have their weight at their lower limit, and all edges not in $P$ have their weight at their upper limit.
The idea is that if we increase the weight of an edge not in $P$, then the length of $P$ does not change, while the $\ell$ might increase, even though the corresponding path might change.
Also if we decrease the weight of an edge in $P$ by some amount $\delta>0$, then the length of $P$ decreases by $\delta$, while $\ell$ might decrease by at most $\delta$.

\begin{theorem}	
	\label{thm:sp}
	Computing the threshold of inclusion of an edge for the $s-t$ shortest path problem is NP-complete.
\end{theorem}

\begin{proof} 
	By Observation~\ref{obs:in-NP}  we know that the problem is in NP.

	The proof of NP-hardness is a reduction from 3-SAT, and is an adaption of the proof given in \cite{gabow_two_1976}.
An instance of 3-SAT consists of $n$ boolean variables $X_1,\ldots,X_n$, and $m$ clauses.
Every clause ${\mathcal{C}}_j$ contains exactly 3 literals, where a literal is either a variable or its negation.
The goal is to find a boolean assignment to the variables which satisfies each clause.
A clause is satisfied if it at least one of its literal is True. 

	Given this 3-SAT instance we construct an instance to the shortest path problem with uncertain edge weights, see Figure~\ref{fig:reduction}.
There will be a vertex for every literal, and for each occurrence of a literal in a clause.
In addition we have the vertices $s,u,v,t$.
The graph has several layers.
Layer $0$ contains vertex $s$.
Layer $j=1,\ldots, m$ contains 3 vertices $u_{1j},u_{2j},u_{3j}$ corresponding to the 3 literals in clause $C_j$.
Layer $m+1$ contains vertex $u$, layer $m+2$ vertex $v$.
Then for every $i=1,\ldots,n$, layer $m+2+i$ contains 2 vertices $v_{0i}$ and $v_{1i}$, corresponding respectively to $\overline{X_i}$ and $X_i$.
Finally layer $m+3+n$ contains vertex $t$.
There are two type of edges.
Solid edges have lower weight 0 and upper weight $1$, and connect all pairs of vertices between two adjacent layers.
The other type of edges are dashed and have weight $0$ (same lower and upper bound).
There is a dashed edge between vertices $u_{kj}$ and $v_{bi}$ if and only if the literal corresponding to $u_{kj}$ is the negation of the literal corresponding to $v_{bi}$.

	\begin{figure}[ht]
		\centering
		\includegraphics[width=\textwidth]{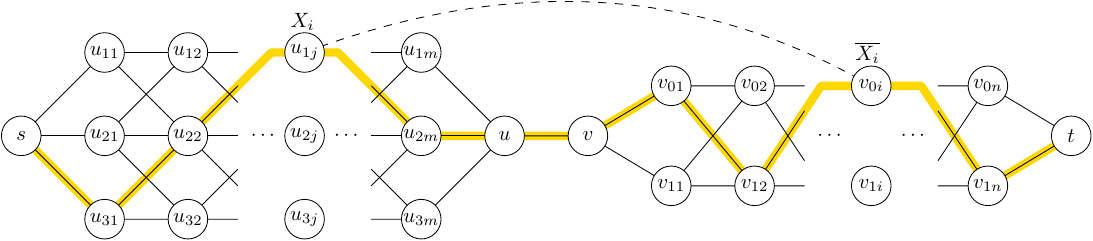}
		\caption{Reduction from 3-SAT. 
		Solid edges have uncertainty intervals $[0,1]$, while dashed edges trivial uncertainty intervals $0$.}
		\label{fig:reduction}
	\end{figure}

	Now we claim that the 3-SAT formula is satisfiable if and only if the threshold of inclusion $\tlow_{uv}$ of the edge $uv$ is at least $-1$.

	For this purpose we introduce the notion of consistency of an $s-t$ path $P$ going through edge $uv$.
We say that $P$ is consistent if it traverses at most one of both endpoints of every dashed edge, and if it traverses every layer exactly once.
We can associate a satisfying assignment to every consistent path.
Formally, let $P$ be a consistent path.
For every $i=1,\ldots,n$ it traverses exactly one of the vertices $v_{0i},v_{1i}$.
If it traverses $v_{bi}$, then we set $X_i=1-b$.
Now each layer $j=1,\ldots,m$ is traversed by $P$ in at least one vertex, say $u_{kj}$, and by consistency this vertex corresponds to a true valued literal.

For one direction of the proof, suppose that the formula is satisfiable.
Let $P$ be a consistent $s-t$ path going through edge $uv$, corresponding to a satisfying assignment.
This path has weight $0$.
Now we consider the shortest $s-t$ path $Q$ not using edge $uv$.
This path has to use at least of the dashed edges, and by consistency of $P$ has to use also at least one solid edge not in $P$.
Hence the weight of $Q$ is at least $1$.

The other direction of the proof is straightforward.
If there is no $s-t$ path $P$ using edge $uv$ which can enforce weight at least $1$, then there is no satisfying assignment.
\end{proof}

\begin{observation}
\label{obs:hardness-path}
We actually prove that it is NP-hard to distinguish whether the threshold of inclusion is $0$ or at most $-1$.
Combining this with Lemma~\ref{lem:binary-search}, we obtain that is hard to decide whether a set is an admissible query.
As a consequence, we get hardness of approximation.
Also, this gap can be made as large a constant as we want by simply scaling the weights.
\end{observation}

For the threshold of exclusion the reduction is much simpler.
\begin{theorem}
	\label{thm:sp-exclusion}
	Computing the threshold of exclusion of an edge for the $s-t$ shortest path problem is NP-complete.
\end{theorem}
\begin{proof}
	The proof is a reduction from the Hamiltonian path problem. Let $H=(V,E)$ be a graph with two vertices $s,t\in V$. The Hamiltonian path problem asks for the existence of a path $P$ from $s$ to $v$, which visits all vertices of $V$ exactly once. This problem is NP-complete.

	Denote by $n$ the number of vertices in $H$. We construct a graph $G$ from $H$. All edges in $H$, which are now also in $G$, have lower weight $\ell_e=1$ and upper weight $h_e=n$. We complete the construction by adding two vertices $u,v$, and edges $su,uv,vt$, all with zero weight (matching lower and upper weight limit). 
	
	Now we claim that such that the threshold of inclusion in $G$ for edge $uv$ is at least $n-1$, if and only if $H$ admits a Hamiltonian path. The shortest $s-t$ path with edge $uv$ has length 0. The shortest $s-t$ path without edge $uv$ has a total length which depends on the edge weights. It can have total length $n-1$ if and only if the set of edges having their weight at the lower limit $1$ forms a Hamiltonian path from $s$ to $t$. This concludes the proof.
\end{proof}

\section{Minimum cost perfect matching}

We complement the previous NP-hardness proof by showing hardness of computing the thresholds for matching.

\begin{theorem}\label{thm:perfect-matching}

	Computing the threshold of exclusion of an edge  in the min cost perfect matching problem is NP-complete already for bipartite graphs.
\end{theorem}
\begin{proof}
	We make a reduction from 3-SAT, similar as in Theorem~\ref{thm:sp}, see Figure~\ref{fig:matching-reduction}.

	For every variable $X_i$, there is a \emph{variable gadget} consisting of a complete bipartite graph, between vertices $p_i, n_i$ on one side and vertices $s_i, v_i$ on the other side. 
    Vertex $p_i$ corresponds to the variable $X_i$ and $n_i$ to its negation $\bar X_i$.
 
	For every clause ${\mathcal C}_j$ there is a \emph{clause gadget} in form of a bipartite graph between vertices $a_j,b_j,c_j$ on one side and vertices $c_{1j},c_{2j},c_{3j}$ on the other side. 
    Vertex $c_{kj}$ corresponds to the $k$-th literal in ${\mathcal C}_j$.

	All edges inside these gadgets have uncertainty intervals of $[0,1]$.
	In addition, there is a single edge between two new vertices $u,v$.
	The construction is completed with additional edges of trivial uncertainty interval $0$ (same lower and upper bound), which are detailed below.
	There is an edge from $v$ to every vertex $v_i$, and from $u$ to every vertex $c_{j}$.
	All vertices $c_{kj}$ are connected to the corresponding literal vertex. Formally if the $k$-th literal in ${\mathcal C}_j$ is $X_i$, then $c_{kj}$ is connected to $p_i$. 
    And if this literal is $\bar X_i$, then $c_{kj}$ is connected to $n_i$.

	\begin{figure}[htp]
		\centering
		\includegraphics[width=10cm]{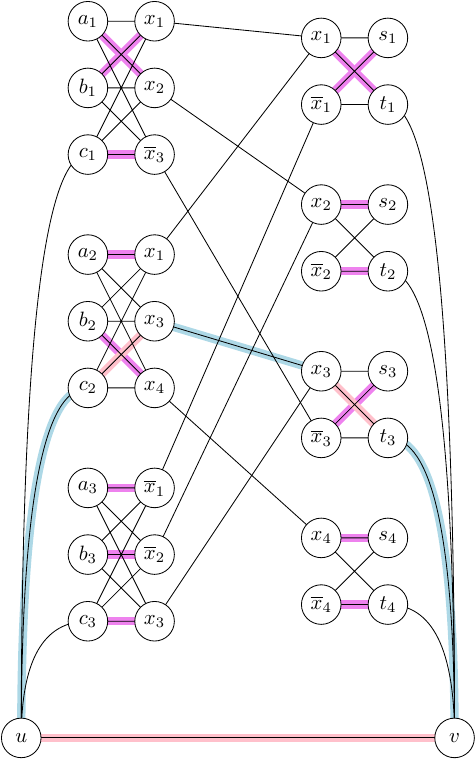}
\caption{The graph obtained by the reduction from the 3-SAT formula consisting of the clauses ${\mathcal C}_1=X_1\vee X_2\vee \overline{X_3},\: {\mathcal C}_2=X_1\vee X_3\vee X_4, \:{\mathcal C}_3=\overline{X_1}\vee \overline{X_2} \vee X_3$. Edges in $M\setminus M'$ are represented by red lines, edges in $M'\setminus M$ by blue lines, and edges in $M\cap M'$ in purple lines. For convenience we replaced vertices of the form $c_{kj},n_i,p_i$ by their corresponding literal.}
\label{fig:matching-reduction}
	\end{figure}

	We claim that the threshold of exclusion of edge $uv$ is $1$ if the formula is satisfiable and $0$ if it is not.

	Consider some edge weights $w$, 
	and let $M^+$ be a min cost perfect matching containing edge $uv$ and $M^-$ a min cost perfect matching not containing edge $uv$. 
    Since $u$ is matched with $v$ in $M^+$, in every clause gadget, the vertex $c_j$ is matched with some vertex $c_{kj}$, and the vertices $a_j, b_j$ are matched to the other two vertices. 
    Similarly, all vertices inside the variable gadgets are matched inside their gadget. 
    These matching edges define a boolean value assignment for the variables, in the sense that $X_i=\text{True}$ if $s_i$ is matched to $p_i$, and $X_i=\text{False}$ if $s_i$ is matched to $n_i$. 
    Also, it selects for every clause one of its literal, by the vertex to which each $c_j$ is matched to. We say that $M^+$ is \defi{satisfying}, if it selects in every clause a literal assigned to true.
	
	The threshold of exclusion is the difference of the costs $M^- - M^+$, maximized over the edge weights $w$. Without loss of generality we can assume that $w_e=0$ for all edges in $e\in M^+$, and $w_e=1$ for all other edges $e\not\in M^+$ and which are inside the gadgets. 
    Remember, edges that are not inside gadgets have weight 0; i.e.,  a matching lower and upper bound of zero.

	We observe that $M^-$ has to match $u$ to some $c_j$ and $v$ to some $v_i$. Also, these have to be the only vertices of the type $c_{j'}$ and $v_{i'}$ which are matched outside of their gadgets. Since $M^-$ is min cost, by the choice of $w$, its matching has to coincide with $M^+$ inside every clause gadget but the $j$-th and inside every variable gadget but the $i$-th. 

    Combining all those observations shows that $M^-$ has zero cost if and only if in $M^+$ one clause selected a literal that it assigned to False.
    Say for example that $M^+$ matches $c_j$ to a vertex $c_{kj}$ corresponding to a literal $X_i$, and matches $s_i$ to $n_i$.
    Then $M^-$ can match $a_j,b_j,s_i$ as in $M^+$ all at cost $0$. However if $M^+$ is satisfying, the cost of $M^-$ is 1. 
    Indeed it can match $a_j,b_j$ is in $M^+$. 
    Assume $(c_{kj},p_i)$ is matched in $M^-$. 
    The case $(c_{kj},n_i)$ is similar.
    Then $s_i$ must be matched to $n_i$. 
    However, since $M^+$ is satisfying by assumption, the cost of this edge is 1.
	This concludes the proof.
\end{proof}

\begin{observation}
\label{obs:hardness-matching}
We actually prove that it is NP-hard to distinguish whether the threshold of exclusion is $1$ or $0$.
Combining this with Lemma~\ref{lem:binary-search} we obtain that it is NP-hard to decide whether a set is an admissible query.
Furthermore, this gap can be made as large a constant as we want by simply scaling the weights.
\end{observation}

The NP-hardness of computing the threshold of inclusion can be shown with a similar construction. Just replace the edge $(u,v)$ by the path $u-u'-v'-v$, where edges $(u,u')$ and $(v',v)$ have both lower and upper weight zero. Now every perfect matching with edge $(u,v)$ in the original graph corresponds to a perfect matching without edge $(u',v')$ in the new graph and vice-versa. Hence computing the threshold of inclusion of edge $(u',v')$ is equivalent to computing the threshold of exclusion of the edge $(u,v)$ in the original graph.

\section{Max weight matching in trees}

We contrast the previous hardness result with a special matching problem for which the thresholds can be computed efficiently.


\begin{theorem}\label{thm:trees}
    The thresholds of inclusion/exclusion can be computed in linear time for a fixed edge $e$ for the maximum weight matching problem on trees.
\end{theorem}
\begin{proof}
We only describe how to compute the threshold of inclusion, as computing the threshold of exclusion follows the same approach. 

Fix an edge $e$ in tree. The edge separates the tree into say
a \emph{left} and a \emph{right} subtree. By $T_u$ we denote the
subtree rooted in some vertex $u$, where $u$ is not necessarily an endpoint of
$e$. The tree $T_u$ consists of all vertices $v$, such that the path from $v$
to $e$ contains vertex $u$.

The threshold of inclusion of edge $e$ is the difference $w (M^-) - w (M^+)$
minimized over weight realizations $w$, with $w (e) = 0$. We can assume
without loss of generality $w (e') = h (e')$ for all $e' \in M^+ \setminus \{
e \}$ and $w (e') = \ell (e')$ for all edges $e' \not\in M^+$.

For an arbitrary weight realization $w$, let $M^+$ be a matching containing
edge $e$ and maximizing total weight. Similarly let $M^-$ be a matching not
containing edge $e$ and maximizing total weight. By breaking ties consistently
between $M^+$ and $M^-$ we obtain that the symmetric difference $M^+ \Delta
M^-$ consists of a single path $P$ containing edge $e$. This path $P$ connects
two vertices $s, t$ where $s$ is in the left subtree and $t$ in the right
subtree. By \emph{neighboring edges} of $P$ we mean edges intersecting $P$
at exactly one endpoint. The other endpoint is called a \emph{neighboring
vertex} of $P$.

The vertices $s$ and $t$ determine in a unique manner the matchings $M^+$ and
$M^-$. Here is how: First the edges along the $s - t$ path belong
alternatively to $M^+$ and to $M^-$. The alternation is defined with respect
to the distance from the edge $e$ in this path. Namely edges $e'$ at odd
distance to $e$ belong to $M^+$ and have weight $h (e')$, while edges at even
distance to $e$ belong to $M^-$ and have weight $\ell (e')$. Neighboring edges
of the $s - t$ path have low weight and do not belong to $M^+$ nor $M^-$.
Finally for every neighboring vertex $u$, we consider a maximum weight
matching $O_u$ in $T_u$ were edges in $O_u$ have high weight and edges in $T_u
\setminus O_u$ have low weight. This sounds as an intricate definition,
because the matching depends on the weight and vice versa, but both can be
computing by bottom up dynamic programming.

Since $M^+$ and $M^-$ coincide outside of the $s - t$ path, the intersection
of $T_u$ with $M^+$ is exactly $O_u$, and so is the intersection of $T_u$ with
$M^-$.

This observation leads to a simple polynomial time procedure for computing the
threshold of inclusion for $e$. Simply loop over all vertices $s$ in the left
subtree and over all vertices $t$ in the right subtree. For each $s - t$ path
compute the matchings $M^+$ and $M^-$ as described in the previous paragraph,
and set the weights of all edges in $M^+$ to their highest value and the
weights of all other edges to their lowest value. At this point we can check
if $M^+$ (respectively $M^-$) is indeed a maximum weight matching under the
condition that it contains $e$ (resp. does not contain $e$). In this case we
call the $s - t$ path a \emph{valid} path. The minimum difference $w (M^-)
- w (M^+)$ over all valid $s - t$ paths is the threshold of inclusion of $e$.

However it is possible to compute the threshold in linear time using dynamical programming, as it is often the case for problems defined on trees.

It is well known from matching theory, that matching $M^+$ has maximum weight, if there is no
augmenting path with respect to $M^+$. Such a path does not contain edge $e$,
by the requirement on $e$. Also such a path must intersect the $s - t$ path by
construction of $M^+$ which is {\emph{locally optimal}} outside of the $s - t$
path. Hence we can verify the validity of the $s - t$ path independently on
its portions in the left subtree and in the right subtree. The same
observation holds for $M^-$. In that sense, an $s - t$ path is valid of the
portion from $s$ to $e$ is valid and the portion from $e$ to $t$ is valid.
These properties are independent. If they hold we say that $s$ and $t$ are
{\emph{valid vertices}}. Now we describe how to enumerate in linear time all
valid vertices $s$ in one of the subtrees.

For two vertices $u, v$, such that $v$ lays on the path from $u$ to $e$, we
define $A_{u v}$ to be the following alternating edge weight sum over the $u -
v$ path.
\begin{itemize}
  \item If $e'$ is at even distance from $e$ and at even distance from $u$, it
  counts with weight $- \ell (e')$.
  
  \item If $e'$ is at even distance from $e$ and at odd distance from $u$, it
  counts with weight $+ \ell (e')$.
  
  \item If $e'$ is at odd distance from $e$ and at even distance from $u$, it
  counts with weight $- h (e') .$
  
  \item If $e'$ is at odd distance from $e$ and at odd distance from $u$, it
  counts with weight $+ h (e')$.
\end{itemize}
These quantities become handy when stating inequalities about paths
alternating with respect to $M^+$ or $M^-$. Note that if $u, v$ are at odd
distance we have $A_{u v} = A_{v u}$ while if they are at even distance we
have $A_{u v} = - A_{v u}$.

First we focus on $O_u$, the maximum weight matching in the subtree rooted at
vertex $u$. In addition we define $O^-_u$ the weight of the maximum weight
matching in this subtree, with the restriction that $u$ is unmatched. It could
be that later in the whole tree $u$ is matched, but not inside the subtree.

At some moment we need to consider the tree $T_u \setminus T_v$ for some
direct descendant $v$ of $u$. We denote by $O^{^-}_{u \setminus v}$ the
maximum weight matching in $T_u \setminus T_v$, with the restriction that $u$
is unmatched. It is simply the sum of $O_{v'}$ over all direct descendants
$v'$ of $u$ but different from $v$. These values can be computed in linear
time using dynamic programming, in leaf to root order of the tree. For example
we have $O_u = \max \{ O^-_u, \max_v h (uv) + O^-_v - O_v \}$.

In addition we define $B_{u \setminus v}$ to be the maximum weight of an
alternating $u - v'$ path in $T_u \setminus T_v$ maximized over $v'$. Edges
$e'$ have weight $- h (e')$ if they belong to the maximum weight matching in
$T_u \setminus T_v$ and have weight $+ \ell (e')$ otherwise.

Before describing the actual algorithm, let's better understand what it means
that $s$ is valid. Consider the path $P$ from $s$ to $e$, and consider a path
$Q$ intersection $P$ in a single portion, namely between two vertices $u$ and
$v$. To fix the notation, assume that $v$ is closer to $e$ than $u$. If $Q$ is
meant to be alternating with respect to $M^+$ or $M^-$, then $u$ and $v$ must
be at odd distance. Let $v'$ be the neighbor of $v$ which is closes to $u$.
$Q$ is not an augmenting path if the following condition hold. It depends on
two cases.
\begin{itemize}
  \item If $u = s$, we must have $B_s + A_{s v} + B_{v \setminus v'} \leqslant
  0$.
  
  \item If $u \neq s$, let $u'$ the neighbor of $u$ which is closes to $s$.
  Then we must have $B_{u \setminus u'} + A_{u v} + B_{v \setminus v'}
  \leqslant 0$.
\end{itemize}
\begin{figure}[h]
  \centerline{\includegraphics[width=10cm]{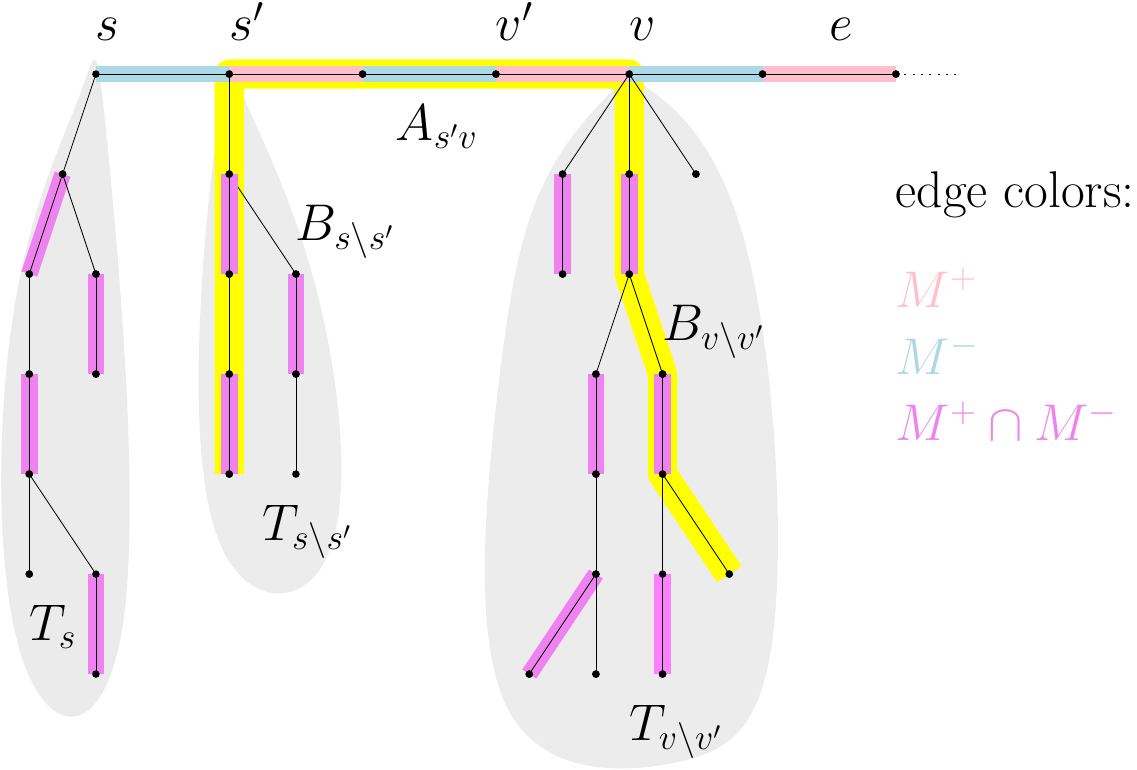}}
  \caption{Technical aspects of verifying the validity of a path}
\end{figure}

While we are exploring in depth first manner the left subtree, we verify these
conditions along the way in the following manner. Let $s$ be the current
vertex of the tree exploration, and $s'$ its ancestor.

If $B_{s' \setminus s} + C_{s'} > 0$ we abort the exploration of this edge
$(s', s)$ as no path containing it is valid. Otherwise we continue the tree
exploration, but before mark $s$ as being a valid vertex in case if $B_s + C_s
> 0$.

Let $u  u'$ be the endpoints of $e$. Let $v $ be a valid vertex which
minimizes $A_{ v u }$ and $v'$ be a valid vertex which minimizes $A_{v' u'}$.
The threshold of inclusion of $e$ is precisely $A_{v u} + A_{v' u'}$.
\end{proof}

\section{Open Questions}

We finish the paper by pointing out some open questions.

Further exploration of the non-adaptive explorable uncertain setting seems an interesting direction.
In particular, to better  identify the boundary between polynomial time tractability and NP-hardness of computing thresholds and finding minimum-cost admissible queries.
Interesting settings not covered in the paper are: Matching in planar graphs, and computing the edit distance between two strings under uncertain edit costs. 
Another intriguing question is whether the tractability of the threshold of inclusion is related to the one of the threshold of exclusion; in the sense that both can be solved in polynomial time or none.

Outside of the setting we study in the paper, it would be interesting to study models that interpolate between being fully adaptive and no adaptivity at all; e.g., models with a fixed number of rounds in which to perform the queries.


\section{Acknowledgments}

We are very grateful to Shahin Kamali, Marcos Kiwi and Claire Mathieu for helpful discussions. 

This work is partially supported by the grants ANR-19-CE48-0016 and  ANR-23-CE48-0010 from the French National Research Agency, the Proyecto Fondo Basal FB210005 CMM and Fondecyt 1221460 and 1231669, ANID.

\printbibliography

\end{document}